\documentclass[twocolumn]{IEEEtran}
\usepackage{cite}
\usepackage{amsmath,epsfig,amssymb,verbatim,amsopn,cite,multirow}
\usepackage{amsthm}
\usepackage{balance}
\usepackage{multirow}
\usepackage[usenames,dvipsnames]{color}
\usepackage[all]{xy}  
\usepackage{url}
\usepackage{amsfonts}
\usepackage{amssymb}
\usepackage{epsfig}
\usepackage{epstopdf}
\usepackage{bm}
\usepackage{balance}
\usepackage{graphicx}
\usepackage{subcaption}
\usepackage{footnote}
\usepackage{cancel}
\usepackage{algorithm,algorithmic}
\usepackage{balance}
\usepackage[margin=14.5mm,top=18.1mm,bottom=25.7mm]{geometry}

\newcommand{\qa}{{\bf a}}
\newcommand{\qb}{{\bf b}}

\newcommand{\qe}{{\bf e}}

\newcommand{\qg}{{\bf g}}
\newcommand{\qh}{{\bf h}}

\newcommand{\qn}{{\bf n}}

\newcommand{\qu}{{\bf u}}
\newcommand{\qv}{{\bf v}}
\newcommand{\qw}{{\bf w}}
\newcommand{\qx}{{\bf x}}
\newcommand{\qy}{{\bf y}}

\newcommand{\qA}{{\bf A}}
\newcommand{\qB}{{\bf B}}

\newcommand{\qH}{{\bf H}}
\newcommand{\qI}{{\bf I}}

\newcommand{\qN}{{\bf N}}

\newcommand{\qY}{{\bf Y}}

\newcommand{\Ex}{\mathbb{E}}
\newcommand{\SSE}{\mathrm{SSE}}
\newcommand{\SE}{\mathrm{SE}}
\newcommand{\des}{\mathrm{des}}
\newcommand{\PZF}{\text{PZF}}
\newcommand{\PMRT}{\text{PMRT}}

\begin{document}
\title{{Resilient Cell-Free Massive MIMO Networks}
\thanks{}}

\author{
    \IEEEauthorblockN{Junbin Yu, Tianyu Lu, Mohammadali Mohammadi, and Michail Matthaiou}
    \IEEEauthorblockA{
        \\ Centre for Wireless Innovation (CWI), Queen’s University Belfast, Belfast, U.K. \\
        E-mail: \{jyu17, t.lu, m.mohammadi, m.matthaiou\}@qub.ac.uk}
        \thanks{This work was supported by the European Research Council (ERC) under the European Union’s Horizon 2020 research and innovation programme (grant agreement No. 101001331) and by a research grant from the Department for the Economy Northern Ireland under the US-Ireland R\&D Partnership Programme.}}

\maketitle

\begin{abstract}
This paper proposes a novel optimization framework for enhancing the security resilience of cell-free massive multiple-input multiple-output (CF-mMIMO) networks with multi-antenna access points (APs) and protective partial zero-forcing (PPZF) under active eavesdropping.
Based on the main principles of absorption, adaptation, and recovery, we formulate a security-aware resilience metric to quantify the system performance during and after a security outage. A multi-user service priority-aware power allocation problem is formulated to minimize the mean squared error (MSE) between real-time and desired security efficiency, thereby enabling a trade-off between the target user's secrecy performance and multi-user quality of service (QoS). To solve this non-convex problem, a security-aware iterative algorithm based on the successive convex approximation (SCA) is employed. The proposed algorithm determines the optimal power allocation strategy by balancing solution quality against recovery time. At each iteration, it evaluates the overall resilience score and selects the strategy that achieves the highest value. Simulation results confirm that the proposed framework significantly improves the resilience of CF-mMIMO networks, allowing flexible adaptation between rapid recovery and high-quality recovery, depending on system requirements.
\end{abstract}


\vspace{-0.5em}
\section{Introduction}
With the development toward sixth-generation (6G) wireless networks, the demand for ultra-reliable, low-latency, and high-capacity communication services is steadily increasing, while application scenarios are becoming more diverse~\cite{matthaiou2021road}. In this context, CF-mMIMO architectures have emerged as a promising solution. CF-mMIMO serves users through a large number of distributed APs and eliminates traditional cellular boundaries. This reduces inter-cell interference, providing more consistent service quality across the entire coverage area \cite{hien,Femenias,mohammadi2024next}.

However, CF-mMIMO networks still face inherent security threats in wireless channels, such as eavesdropping, interference, and spoofing, especially in critical mission scenarios. Although traditional cryptographic methods can provide protection at the logical layer, they typically also incur significant computational overhead and latency, especially in dense Internet of Things (IoT) and edge computing environments. As a complementary method, physical layer security (PLS) has gained increasing attention for ensuring data confidentiality by utilizing the randomness of wireless channels and beam design \cite{junbin,ASAN}. Nevertheless, in practical deployments, secure transmission in CF-mMIMO networks can be severely affected by dynamic and unpredictable factors (including physical blockages, AP failures, and malicious attacks), which may cause the service quality to drop
\cite{yasseen2,Tubail,Hoang:TWCOM:2018,yasseen1}.

These challenges underscore the need for secure, resilient communication capable of withstanding malicious attacks while maintaining acceptable QoS under adverse conditions \cite{PLS6G}. In this work, we adopt a resilience framework to assess the system’s ability to absorb disruptions, adapt to changing conditions, and recover within a defined time. Unlike robustness—which preserves performance under predefined conditions—resilience emphasizes adaptive response and recovery during unexpected interruptions \cite{Reifert1}. We extend this framework to the security domain by developing a security-aware resilience design for CF-mMIMO networks under active eavesdropping.

In this work, we establish the first comprehensive framework for resilience-aware PLS in CF-mMIMO systems. Our main contributions are as follows:
\begin{itemize}
\item{We first incorporate the concept of resilience into the framework of PLS in CF-mMIMO systems. Based on the main principles of absorption, adaptation, and recovery, the proposed framework, unlike traditional robustness, quantifies and optimizes the system's dynamic recovery capabilities in the face of unforeseen security disruptions.}
\item{Under the actual constraints of CF-mMIMO systems, we formulate a novel optimization problem centered around service priorities. The adaptation metric of our resilience model is defined via a weighted MSE objective function. The weights $\omega_i$ within this objective serve as the service priorities, enabling a flexible trade-off between the secrecy performance for the targeted user and the QoS for all users, providing a new paradigm for scenarios with multi-objective service level agreements (SLAs).}
\item{We introduce a resilience-aware iterative algorithm that incorporates a second-level resilience priority weights $\lambda_i$. Instead of merely seeking a converged optimal solution, the algorithm evaluates a candidate solution at each iteration by trading off between quality and recovery time, ultimately selecting the strategy with the highest overall resilience score for deployment.}
\end{itemize}

\textit{Notation:} The boldface letters used in this paper represent vectors, while the uppercase letters denote matrices. The terms $(\cdot)^\mathrm{T},(\cdot)^\mathrm{H}$ denote the transpose operator, conjugate transpose operator; $\qI_M$ denotes the $M\times M$ identity matrix; $\|\cdot\|$ denotes the Euclidean norm of a vector or the Frobenius norm of a matrix, while $|\cdot|$ represents the absolute value of a scalar.  A zero-mean circular-symmetric complex Gaussian variable having variance $\sigma^2$ is denoted by $\mathcal{CN}(0,\sigma^2)$. The operator $\mathrm{diag}(\qA)$ denotes a vector consisting of the diagonal elements of matrix $\qA$.  Finally, $\mathbb{E}\{\cdot\}$ denotes the statistical expectation.

\section{System Model}
\begin{figure}[t]
\centering
\includegraphics[width=2.6 in]{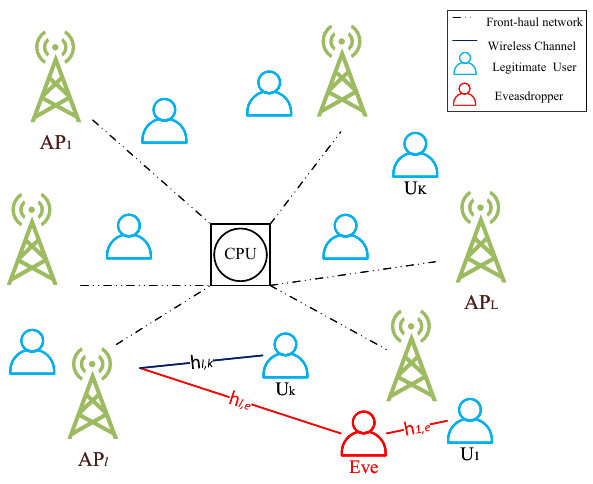}
\caption{Illustration of secure CF-mMIMO network.}
\label{fig1}
\end{figure}

We consider a CF‑mMIMO network illustrated in Fig.\,\ref{fig1}, which operates in the presence of an eavesdropper (Eve). In particular, $L$ APs with $M$ transmit antennas are linked to a central processing unit (CPU) via a front-haul network. All APs simultaneously send data to $K$ users with a single antenna, while a single antenna Eve passively eavesdrops on the data. For simplicity of notation, we define the sets $\mathcal{K}\triangleq\{1,\ldots,K\}$, and $\mathcal{L}\triangleq\{1,\ldots,L\}$ to represent the sets of the users, and APs, respectively. We assume that the CF-mMIMO network utilizes a time-division duplex (TDD) protocol, with each coherence interval comprising three sequential phases: (1) uplink training for channel estimation; (2) uplink data transmission; (3) downlink data transmission. In the uplink training phase, users send pilot sequences to APs, each of which estimates the channel for all users. The resulting channel estimates are used to beamform the transmitted signals in the downlink and to detect the signals transmitted from the users in the uplink. 
In the downlink, only data symbols are transmitted; no pilots are sent to estimate the users’ effective channel gains. Instead, the system relies on channel hardening, which ensures that these gains converge to a nearly deterministic expected value.


\vspace{-0.5em}
\subsection{Channel Model}
We assume that all APs are connected to the CPU via perfect fronthaul links that provide sufficient capacity, and that the channel gains are identical in the uplink and downlink due to TDD reciprocity. The channel between the $l$-th AP and the $k$-th user (denoted $U_k$), or Eve, is represented by an $N$-dimensional vector capturing both small-scale and large-scale fading effects, and is modeled as
\vspace{-0.2em}
\begin{equation}
    \qh_{l,\xi}=\sqrt{\beta_{l,\xi}}\qg_{l,\xi},
\end{equation}
where $\beta_{l,\xi}$ and $\qg_{l,\xi}\in\mathbb{C}^{M\times1}$, for $l\in\mathcal{L}=\{1, \dots, L\}$, $\xi \in \{k, e\}$, $k\in\mathcal{K}$, represent the large-scale fading coefficient and small-scale fading vector, respectively.\footnote{In the remainder of this paper, $e$ denotes the parameters corresponding to Eve.} Each element of $\mathbf{g}_{l,\xi}$ is modeled as an independent and identically distributed complex Gaussian random variable, i.e., $\qg_{l,\xi} \sim \mathcal{CN}(0,\qI_M)$.

\vspace{-0.5em}
\subsection{Uplink Training and Active Eavesdropping}
During uplink training, each user $U_k$ simultaneously transmits a unique orthonormal pilot sequence $\boldsymbol{\phi}_k \in \mathbb{C}^{{\tau_p}\times 1}$ of length $\tau_p\ge K$, $\boldsymbol{\phi}_k^\mathrm{H}\boldsymbol{\phi}_k=1$ and $\boldsymbol{\phi}_k^\mathrm{H}\boldsymbol{\phi}_{k'}=0$, ${k'}\ne k$. We consider an active Eve who performs a pilot contamination attack on $U_1$ by transmitting the same pilot, i.e., $\boldsymbol{\phi}_e=\boldsymbol{\phi}_1$. The signal received at the $l$-th AP is
\begin{equation}
\qY_{l} = \sum\nolimits_{k\in \mathcal{K}} \sqrt{\tau_p p_k} \qh_{l,k} \boldsymbol{\phi}_k^{\mathrm{H}} + \sqrt{\tau_p p_e} \qh_{l,e} \boldsymbol{\phi}_{1}^{\mathrm{H}} + \qN_{l},
\end{equation}
where $p_k \triangleq P_k/N_0$ and $p_e \triangleq P_e/N_0$ denote the uplink pilot powers normalized by the noise power $N_0$ for user $U_k$ and Eve, respectively. Moreover,  $\qN_l\in\mathbb{C}^{M\times\tau_{p}}$ is the additive Gaussian white noise (AWGN) matrix with i.i.d. $\mathcal{CN}(0,1)$ elements.

In order to estimate the channel $\qh_{l,k}$, the $l$-th AP received signal $\qY_l$ is projected onto the pilot sequence $\boldsymbol{\phi}_k$ as
$$\qy_{l,k}=\!\qY_l \boldsymbol{\phi}_k=\sqrt{\tau_p p_k} \qh_{l,k} + \delta_k\sqrt{\tau_p p_e} \qh_{l,e} + \qn_{l,k},$$
where $\delta_k$ equals 1 when $k=1$ and 0 otherwise. Moreover, $\qn_{l,k} \triangleq \qN_l \boldsymbol{\phi}_k$ is the effective noise vector with $\mathcal{CN}(\mathbf{0},\qI_M)$ distribution. Then, the $l$-th AP performs minimum mean square error (MMSE) estimation \cite{LPZF}. The channel estimate $\hat{\qh}_{l,k}$ can be obtained as $\hat{\qh}_{l,k} =\frac{\sqrt{\tau_p p_k} \beta_{l,k}}{\tau_p p_k \beta_{l,k} + \delta_k\tau_p p_e \beta_{l,e} + 1} \qy_{l,k}$. Therefore, it can be shown that $\hat{\qh}_{l,k} \sim \mathcal{CN}(\mathbf{0}, \gamma_{l,k}\qI_M)$ with $\gamma_{l,k}=\frac{\tau_p p_k \beta_{l,k}^2}{\tau_p p_k \beta_{l,k} + \delta_k\tau_p p_e \beta_{l,e} + 1}$. Moreover, the channel estimation error is denoted by $\tilde{\qh}_{l,k} = \qh_{l,k} - \hat{\qh}_{l,k}$, which is statistically independent of the estimate $\hat{\qh}_{l,k}$, and $\tilde{\qh}_{l,k}\sim\mathcal{CN}(\mathbf{0},(\beta_{l,k} - \gamma_{l,k})\qI_M)$~\cite{LPZF}.

\vspace{-0.5em}
\subsection{Downlink Data Transmission}
To counter the active Eve, we introduce artificial noise (AN) in this section. The AN vector is typically in the null space of the legitimate user's channel, i.e., the AN is orthogonal to the user's channel vectors, thus minimizing the power of the AN signal when it reaches the legitimate user and avoiding interference to it \cite{ASAN}.

The total transmit signal vector at the $l$-th AP is expressed as
\begin{equation}
    \qx_l = \sum\nolimits_{k\in \mathcal{K}}\sqrt{\rho_{l,k}}\qw_{l,k}s_k + \sqrt{\rho_{\mathtt{AN},l}}\qv_l n_{\mathtt{AN},l},
\end{equation}
where $\rho_{l,k}$ is the allocated power coefficient, $\qw_{l,k} \in \mathbb{C}^{M \times 1}$ is the corresponding precoding vector, and $s_k$ is the data symbol for $U_k$ with $\Ex\{|s_k|^2\}=1$. The AN component consists of the AN power $\rho_{\mathtt{AN},l}$, a random AN symbol $n_{\mathtt{AN},l} \sim \mathcal{CN}(0,1)$, and a normalized AN beamforming vector $\qv_l \in \mathbb{C}^{M \times 1}$ with $\Ex\{\|\qv_l\|^2\}=1$.

Thus, the signal received at the $U_k$ and Eve is given by
\begin{align}
      y_\xi &=\sum\nolimits_{l\in\mathcal{L}} \qh^\mathrm{H}_{l,\xi}\qx_l \notag\\
      &= \sum_{l\in\mathcal{L}} \sum_{t\in\mathcal{K}} \sqrt{\rho_{l,t}} \qh^\mathrm{H}_{l,\xi}\qw_{l,t} s_t + \sum_{l\in\mathcal{L}} \sqrt{\rho_{\mathtt{AN},l}} \qh^\mathrm{H}_{l,\xi} \qv_l n_{\mathtt{AN},l} + n_\xi,  
\end{align}
where $n_\xi\sim\mathcal{CN}(0,1)$ denotes the AWGN vector received at the $U_k$ or Eve.

\vspace{-0.5em}
\subsection{Downlink Precoding and AN Design}
We adopt the PPZF precoding strategy from \cite{LPZF}. For each AP, users are divided into a strong user set $\mathcal{S}_l$ and a weak user set $\mathcal{W}_l$, based on their large-scale fading gains, and $|\mathcal{S}_l|+|\mathcal{W}_l|=K$. The large-scale fading coefficient and the channel estimation quality are captured by $\beta_{l,k}$ and $\gamma_{l,k}$, respectively, where $\Ex\{\|\qh_{l,k}\|^2\} = M\beta_{l,k}$ and $\Ex\{\|\hat{\qh}_{l,k}\|^2\} = M\gamma_{l,k}$.

The data precoding vector $\qw_{l,k}$ depends on whether $U_k$ is in $\mathcal{S}_l$ or $\mathcal{W}_l$. To protect strong users from interference, both the signals for weak users and the AN are transmitted in the null space of the strong users' estimated channels.

Then, we divide all the APs into two sets of indices: $\mathcal{Z}_k\triangleq\{l:k\in\mathcal{S}_l,l\in\mathcal{L}\}$ and $\mathcal{M}_k\triangleq\{l:k\in\mathcal{W}_l,l\in\mathcal{L}\}$ with $|\mathcal{Z}_k|+|\mathcal{M}_k|=L$, and $\mathcal{Z}_k\cap\mathcal{M}_k=\varnothing$.
\begin{itemize}
    \item \textbf{PZF for Strong Users ($l\in\mathcal{Z}_k, k \in \mathcal{S}_l$)}: The precoding vector $\qw_{l,k} = \qw_{l,k}^{\PZF}$ is designed to nullify interference among strong users. Following \cite{yasseen1}, it is given by $\qw_{l,k}^{\PZF} = \frac{\tilde{\qw}_{l,k}^{\PZF}}{\sqrt{\Ex\{\|\tilde{\qw}_{l,k}^{\PZF}\|^2\}}} = \tilde{\qw}_{l,k}^{\PZF} \sqrt{(M-|\mathcal{S}_l|)\gamma_{l,k}}$,
    where  $\tilde{\qw}_{l,k}^{\PZF} = \hat{\qH}_{l, \mathcal{S}_l} (\hat{\qH}_{l, \mathcal{S}_l}^H \hat{\qH}_{l, \mathcal{S}_l})^{-1} \qe_k$ ($\qe_k$ is the $k$-th basis vector in $\mathbb C^{|\mathcal S_l|\times 1}$), while $\hat{\qH}_{l,\mathcal{S}_l}=[\hat{\qh}_{l,k}: k\in\mathcal{S}_l]\in\mathbb C^{M\times|\mathcal S_l|}$ is the estimated channel matrix for strong users at the $l$-th AP.

    \item \textbf{PMRT for Weak Users ($l\in\mathcal{M}_j, j \in \mathcal{W}_l$)}: The precoding vector $\qw_{l,j} = \qw_{l,j}^{\PMRT}$ is projected into the null space of strong users' channels via the projection matrix $\qB_l$: $\qw_{l,j}^{\PMRT} =\frac{\qB_l \hat{\qh}_{l,j}}{\sqrt{\Ex\{\|\qB_l \hat{\qh}_{l,j}\|^2\}}}=\frac{\qB_l \hat{\qh}_{l,j}}{\sqrt{(M - |\mathcal{S}_l|)\gamma_{l,j}}},$
    where $\qB_l =\qI_M - \hat{\qH}_{l, \mathcal{S}_l}(\hat{\qH}_{l, \mathcal{S}_l}^\mathrm{H} \hat{\qH}_{l, \mathcal{S}_l})^{-1} \hat{\qH}_{l, \mathcal{S}_l}^\mathrm{H} $.

    \item \textbf{AN Vector}: The AN vector is also designed within the null space projected by $\qB_l$ to avoid interfering with strong users, ${\qv}_l = \frac{\qB_l \qa_l}{\sqrt{\Ex\{\|\qB_l \qa_l\|^2\}}}=\frac{\qB_l \qa_l}{\sqrt{M-|\mathcal{S}_l|}},$
where $\qa_l \in \mathbb{C}^{M \times 1}$ is an isotropic random vector and $\qa_l \sim \mathcal{CN}(\mathbf{0},\qI_M)$. Notably, when $l\in\mathcal{Z}_1$ ($U_1\in\mathcal{S}_l$), the projection places the AN in the null space of Eve's estimated channel $\hat{\qh}_{l,e}$, which significantly suppresses leakage. The remaining leakage is only due to the channel estimation error, i.e., $\Ex\{| \qh_{l,e}^\mathrm{H}\qv_l |^2\}=\beta_{l,e}-\gamma_{l,e}$. For $l\in\mathcal{M}_1$, the leakage remains $\beta_{l,e}$, i.e., AN is effective.
\end{itemize}

The total transmit signal from the $l$-th AP to user $U_k$ is $\qx_l=\sum_{k\in\mathcal{S}_l}\sqrt{\rho_{l,k}}\qw_{l,k}^\mathtt{PZF}s_k+\sum_{i\in\mathcal{W}_l}\sqrt{\rho_{l,i}}\qw_{l,i}^\mathtt{PMRT}s_i+\sqrt{\rho_{\mathtt{AN},l}}\qv_ln_{\mathtt{AN},l}.$
Based on previous work \cite{yasseen1}, the additional term introduced by AN can be specifically calculated as $\Ex\Big\{\Big|\sum\nolimits_{l\in\mathcal{L}}\sqrt{\rho_{\mathtt{AN},l}}\qh^{H}_{l,\xi}\qv_{l}n_{\mathtt{AN},l}\Big|^2\Big\} =\sum\nolimits_{l\in\mathcal{L}}\rho_{\mathtt{AN},l}(\beta_{l,\xi}-\delta_{l,k}\gamma_{l,\xi})$,
where $\delta_{l,k}\triangleq1$ if $U_k \in \mathcal{S}_l$ and $\delta_{l,k}=0$ otherwise. Thus, the signal-to-interference-plus-noise ratio (SINR) of $U_k$ can be expressed as follows:
\begin{align}
       &\mathrm{SINR}_k \! =\! \frac{
        \Big( \sum\nolimits_{l\in\mathcal{L}} \sqrt{(M - |\mathcal{S}_l|) \rho_{l,k}\gamma_{l,k}}  \Big)^2
    }{
        \sum\limits_{t\in\mathcal{K}}\! \sum\limits_{l\in\mathcal{L}} \!\rho_{l,t} (\beta_{l,k} - \delta_{l,k} \gamma_{l,k}) 
        \!+\! \!\!\sum\limits_{l\in\mathcal{L}}\! \rho_{\mathtt{AN},l} (\beta_{l,k} \!-\! \delta_{l,k} \gamma_{l,k})\!\!+\!1
    }.\notag
\end{align}

Moreover, the SINR at the Eve when targeting $U_1$ is given by
\begin{align}
        &\mathrm{SINR}_e^{(1)} \\
    &= \frac{
        \Big( \sum\limits_{l\in\mathcal{L}} \sqrt{\rho_{l,1} (M - |\mathcal{S}_l|) \gamma_{l,e}} \Big)^2 
        \!+ \sum\limits_{l\in\mathcal{L}} \rho_{l,1} \beta_{l,e} \!-\!\sum\limits_{l\in\mathcal{Z}_1} \rho_{l,1}\gamma_{l,e}
    }{
        \sum\limits_{t=2}^K \sum\limits_{l\in\mathcal{L}} \rho_{l,t} (\beta_{l,e} - \delta_{l,1}\gamma_{l,e}) 
        + \sum\limits_{l\in\mathcal{L}} \rho_{\mathtt{AN},l} (\beta_{l,e}- \delta_{l,1}\gamma_{l,e}) 
        \!+\! 1
    }, \notag
\end{align}
where $\gamma_{l,e}=\frac{\tau_pp_e\beta_{l,e}^2}{\tau_pp_1\beta_{l,1}+\tau_pp_e\beta_{l,e}+1}$. 

Finally, the secrecy spectral efficiency (SSE) for $U_1$ is given by: 
\begin{align}
    \SSE_{1} &=\!\!\left[\SE_1\!-\! \SE_e^{(1)}\right]^+
\end{align}
where $[x]^+ = \max\{0,x\}$, $\SE_{k}\triangleq\log_2(1\!+\!\text{SINR}_k)$, $\forall k\in\mathcal{K}$, and $\SE_e^{(1)}\triangleq\log_2(1\!+\!\text{SINR}_e^{(1)})$.

\section{Problem Formulation}
For the PPZF precoding scheme, our objective is to optimize the power allocation coefficients $\rho_{l,k}$ and $\rho_{\mathtt{AN},l}$. To model scenarios with heterogeneous service priorities, specifically to protect a target user $U_1$ while maintaining a certain QoS for other users, we introduce the user-level weights $\omega_1$ and $\omega_2$ to adjust the service optimization priority. The optimization problem is formulated to minimize the weighted MSE between the real-time and desired performance metrics:
\begin{subequations}
    \begin{align}
\min_{\rho_{l,k}, \rho_{\mathtt{AN},l}} & \Psi_{\omega} \triangleq\omega_1\left| \frac{\SSE_{1}}{\SSE_{1}^{\des}}\!- 1 \right|^{2}
\!+\!\frac{\omega_2}{K-1} \sum_{k \in \mathcal{K}\setminus \{1\}} \left| \frac{\SE_{k}}{\SE_k^{\des}} - 1 \right|^{2},
\label{Psi_omega}  \\
        \text{s.t.} \quad 
        & \SE_{k} \ge \SE^{\min}_k, \quad \forall k \in \mathcal{K}, \label{eq:SE_min} \\
        & \sum\nolimits_{k\in\mathcal{K}} \rho_{l,k} + \rho_{\mathtt{AN},l} \le P_{\max,l}, \quad \forall l\in \mathcal{L}, \label{eq:rho_power}  \\
        & \rho_{l,k} \ge 0, \quad \rho_{\mathtt{AN},l} \ge 0, \quad \forall l\in \mathcal{L},k\in \mathcal{K}.
    \end{align}\label{P1}
\end{subequations}
The constraint (\ref{P1}b) ensures a minimum rate $\SE^{\min}_k$  for each user $U_k$, thereby satisfying the basic QoS requirement. Constraint (\ref{P1}c) is the maximum transmit power limit for each AP. Constraint (\ref{P1}d) ensures that all power coefficients are non-negative. 

Problem~\eqref{P1} is non-convex because of the fractional SINR expressions and the inherent complexity of the objective function.  To handle the non-convex objective function, we first decouple the difference in the objective and address the non-convex SINR terms. To this end, we introduce the auxiliary variables $\tau_k$ and $\eta_1$ to represent, respectively, the lower bound of $\SE_k$ and the upper bound of $\SE_e^{(1)}$. Then, we define $\tau_k \le \SE_k, \eta_1 \ge \SE_e^{(1)}$, and set $\zeta_1 \triangleq [\tau_1 - \eta_1]^+ =\max\{0,\tau_1-\eta_1\}\ge0$ as the $\SSE_1$ surrogate. Moreover, to convexify the per-AP power constraint (\ref{P1}c), we define $u_{l,k}=\sqrt{\rho_{l,k}}$ and $u_{\mathtt{AN},l}=\sqrt{\rho_{\mathtt{AN},l}}$,and set $\mathbf{u}=[\mathbf{u}_k,\mathbf{u}_{\mathtt{AN}}]^T\in\mathbb{R}^{(KL+L)\times 1}$. The original problem (\ref{P1}) can be reformulated as follows:
\begin{subequations}
\label{P2}
\begin{align}
\min_{\qu, \boldsymbol{\tau}, \eta_1, \zeta_1} ~ &  
\Psi_\omega = \omega_1\left| \frac{\zeta_1}{\SSE_1^{\des}} - 1 \right|^2\!+\!\frac{\omega_2}{K-1}\sum_{k\ne1}\left|\frac{\tau_k}{\SE_k^{\des}}-1\right|^2 \label{obj_omega} \\
\text{s.t.} \quad
& \log_2(1 + \mathrm{SINR}_k(\boldsymbol{\qu})) \ge \tau_k, \quad \forall k\in\mathcal{K}, \label{eq:SE_tau} \\
& \log_2(1 + \mathrm{SINR}_e^{(1)}(\boldsymbol{\qu})) \le \eta_1, \label{eq:SE_eta}\\
& \tau_k \ge \SE^{\min}_k,\quad \forall k\in\mathcal{K},\label{eq:tau_min}\\
&\tau_1 - \eta_1 \ge \zeta_1 \ge 0,  \label{eq:SSE_zeta}\\
& \sum\nolimits_{k\in\mathcal{K}} u_{l,k}^2 + u_{\mathtt{AN},l}^2 \le P_{\max,l}, \quad \forall l\in\mathcal{L}\label{eq:power_u}\\
& u_{l,k} \ge 0, \quad u_{\mathtt{AN},l} \ge 0 \quad\forall k, l.\label{eq:u0}
\end{align}\label{P2}
\end{subequations}
\!\!The optimization variables are now $(\qu,\boldsymbol{\tau},\eta_1,{\zeta_1})$. This optimization problem remains non-convex due to the non-convex nature of constraints \eqref{eq:SE_tau} and \eqref{eq:SE_eta}, which will be addressed using the proposed SCA-based iterative algorithm.

To address the non-convex constraint \eqref{eq:SE_tau}, the SINR of $U_k$ can now be written as:
\begin{equation}
    \mathrm{SINR}_k(\qu) =\frac{(\qa_k^\mathrm{T} \qu_k)^2}{\varphi_k(\qu)},
    \label{SINR_k}
\end{equation}
where 
\begin{subequations}
    \begin{align}
        \qa_k&= \left[\sqrt{(M - |\mathcal{S}_1|)\gamma_{1,k}}, \dots, \sqrt{(M - |\mathcal{S}_L|)\gamma_{L,k}} \right],\\
        \qA_{k,k}&=\mathrm{diag}\left(\sqrt{\beta_{1,k}-\delta_{1,k}\gamma_{1,k}}, \dots, \sqrt{\beta_{L,k}-\delta_{L,k}\gamma_{L,k}}\right),\\
        \varphi_k(\qu) &=\sum_{t\in\mathcal{K}}\|\qA_{k,k}\qu_t\|^2+\|\qA_{k,k}\qu_{\mathtt{AN}}\|^2+1
    \nonumber\\
    &\hspace{-2em}=\sum_{t\in\mathcal{K}} \sum_{l\in\mathcal{L}} (\beta_{l,k} - \delta_{l,k}\gamma_{l,k}) u_{l,t}^2 
    \!+\!\sum_{l\in\mathcal{L}} (\beta_{l,k} - \delta_{l,k}\gamma_{l,k}) u_{\mathtt{AN},l}^2\!+\!1.
    \end{align}
\end{subequations}
 
At the $(n+1)$-th iteration, using the first-order global lower bound of the convex function $x^2/\varphi$ for $\varphi>0$ around $(x_k^{(n)},\varphi_k^{(n)})$, where $x_k^{(n)}\triangleq\qa_k^\mathrm{T}\qu_k^{(n)}$ and $\varphi_k^{(n)}\triangleq\varphi_k(\qu^{(n)})$, constraint \eqref{eq:SE_tau} can be approximated by the following convex form: 
\begin{equation} \label{eq:user_rate}
\tau_k \le \log_2(1 + \mathrm{SINR}_k^{\mathrm{lb}}(\mathbf{u}, \mathbf{u}^{(n)})), \quad \forall k \in \mathcal{K}, 
\end{equation} 
where
\begin{equation} 
\mathrm{SINR}_k^{\mathrm{lb}}(\mathbf{u}, \mathbf{u}^{(n)}) \triangleq \frac{2x_k^{(n)}}{\varphi_k^{(n)}} x_k(\mathbf{u}) - \frac{(x_k^{(n)})^2}{(\varphi_k^{(n)})^2} \varphi_k(\mathbf{u}). 
\end{equation}

Similarly, to address the non-convex constraint \eqref{eq:SE_eta}, the SINR at the Eve targeting user $U_1$ can be expressed as a convex quadratic ratio function of $\mathbf{u}$ as follows
\begin{equation}
    \mathrm{SINR}_e^{(1)}(\qu) = \frac{\mathrm{f_N}_e(\qu)}{\mathrm{f_D}_e(\qu)}=\frac{(\qb^\mathrm{T}_e\qu_1)^2+\|\qB_e\qu_1\|^2}{\sum_{t=2}^K\|\qB_e\qu_t\|^2+\|\qB_e\qu_{\mathtt{AN}}\|^2+1},
    \label{SINR_e}
\end{equation}
where 
\begin{subequations}
  \begin{align}
   & \qb_e=\left[\sqrt{(M-|\mathcal{S}_1|)\gamma_{1,e}}\, \dots,\sqrt{(M-|\mathcal{S}_L|)\gamma_{L,e}}\right], \\
&\qB_e=\mathrm{diag}\left(\sqrt{\beta_{1,e}-\delta_{1,1}\gamma_{1,e}}, \dots,\sqrt{\beta_{L,e}-\delta_{L,1}\gamma_{L,e}}\right).
\end{align}  
\end{subequations}

Then, we introduce an auxiliary variable $\gamma_e\ge0$ to serve as an upper bound for the SINR term, i.e., $\mathrm{SINR}_e^{(1)}(\qu) \le \gamma_e$.
The upper-bounding condition then becomes
\begin{equation}~\label{eq:boundgamae}
    \gamma_e \ge \frac{\mathrm{f_N}_e(\qu)}{\mathrm{f}_{\mathrm{D}e}^\mathrm{lb}(\qu, \qu^{(n)})} = \frac{(\qb^\mathrm{T}_e\qu_1)^2+\|\qB_e\qu_1\|^2}{\mathrm{f}_{\mathrm{D}e}^\mathrm{lb}(\qu, \qu^{(n)})}.
\end{equation}
Equivalently, denoting $\mathbf{x}_e\triangleq[\qb_e^\mathrm{T} \qu_1; \qB_e \qu_1] \in \mathbb{R}^{(L+1)\times 1}$, we can rewrite~\eqref{eq:boundgamae} as
\begin{equation}~\label{eq:quad}
  \|\mathbf{x}_e\|^2 \le \gamma_e\,\mathrm{f}_{\mathrm{D}e}^\mathrm{lb}(\qu,\qu^{(n)}). 
\end{equation}
Since the left-hand side of~\eqref{eq:quad} is quadratic, this inequality is non-convex. However, it can be reformulated as the following convex second-order cone (SOC) constraint:
\begin{equation}
\label{eq:eve_soc_rate}
\left\| 
\begin{bmatrix}
2\qx_e \\
\gamma_e - \mathrm{f}_{\mathrm{D}e}^\mathrm{lb}(\qu, \qu^{(n)})
\end{bmatrix}
\right\| \le \gamma_e + \mathrm{f}_{\mathrm{D}e}^\mathrm{lb}(\qu, \qu^{(n)}).
\end{equation}

Then, we consider the SINR variable $\gamma_e$ in the $\SE_e$. To create a tight concave lower bound for the function $\log_2(1+\gamma_e)$, we use its first-order Taylor expansion around the value $\gamma_e^{(n)}$ obtained from the previous iteration $n\ge0$:
\begin{equation}
\label{eq:eve_rate}
\eta_1 \ge \log_2(1+\gamma_e) \ge \log_2(1+\gamma_e^{(n)}) + \frac{\gamma_e - \gamma_e^{(n)}}{(1+\gamma_e^{(n)})\ln 2},
\end{equation}
where $\gamma_e^{(n)}= \mathrm{SINR}_e^{(1)}(\qu^{(n)})$.
The SOC constraint (\ref{eq:eve_soc_rate}) and the linear constraint (\ref{eq:eve_rate}) constitute a convex approximation for (\ref{P2}c).

At each inner iteration $(n+1)$ of the SCA algorithm, we solve the following convex subproblem, which is an SOCP problem:
\begin{subequations}
\label{eq:SOCP}
\begin{align}
    \min_{\qu, \boldsymbol{\tau}, \eta_1, \zeta_1, \gamma_e} \quad & \Psi_\omega \\
    \text{s.t.} \quad 
    & (\ref{P2}\mathrm{d}), (\ref{P2}\mathrm{e}), (\ref{P2}\mathrm{f}), (\ref{P2}\mathrm{g}), (\ref{eq:user_rate}), (\ref{eq:eve_soc_rate}), (\ref{eq:eve_rate}).
\end{align}
\end{subequations}

We iteratively solve the SOCP problem \eqref{eq:SOCP} to update the variables. Specifically, after obtaining $(\qu^*, \boldsymbol{\tau}^*, \eta_1^*)$ of the problem, we update the starting point $(\qu^{(n+1)}, \boldsymbol{\tau}^{(n+1)}, \eta_1^{(n+1)})$ in the next iteration until convergence, thereby finding a locally optimal power allocation for the original problem (\ref{P1}). 

\textbf{Complexity analysis:} The computational complexity of each iteration of~\eqref{eq:SOCP} is dominated by solving the associated SOCP, leading to an overall complexity of $\mathcal{O}\!\left(I (KL)^3\right)$, where $I$ denotes the number of iterations.

\section{Secure Resilience-Aware Framework}
\begin{figure}[t]
\centering
\includegraphics[width=3.2 in]{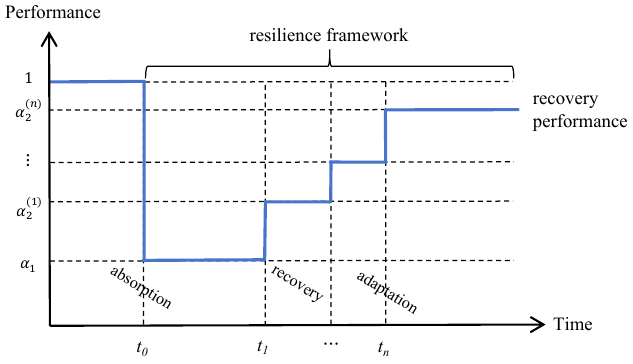}
\caption{Illustration of the security
resilience framework.}
\label{resilience}
\end{figure}
After designing the SCA-based iterative algorithm to find local optima, we introduce the secure resilience-aware framework that addresses the dynamic system recovery problem. Unlike traditional convergence optimization, it evaluates and selects power allocation strategies by simultaneously considering solution quality and optimization time. As illustrated in Fig.\,\ref{resilience}, we introduce a resilience metric to balance between the security performance and network efficiency. It can be divided into three parts following the resilience framework in \cite{Reifert1}:
\subsubsection{Absorption} In the part of \textit{absorption}, $t_0$ is the time point when the overall performance deteriorates. Considering the time-varying $\SSE_{1}(t)$, $\SE_k(t)$ and the constant desired $\SSE^\des_ 1$, $\SE^\des_ k$, the absorption can be expressed as:
\begin{equation}
    \alpha_{\mathtt{abs}}=1-\Psi_\omega(t_0),
    \label{abs}
\end{equation}
where $\Psi_\omega(\cdot)$ is the weighted gap objective (defined in \eqref{Psi_omega}).
\subsubsection{Adaptation}In the part of \textit{adaptation}, the resilience controller triggers to improve security performance using available resources. The metric is defined as the ratio of the SSE to the desired SSE at the time $t_n$, where $t_n$ is the time at which the automatic recovery run ends, of reaching the recovery state. The adaptation can be expressed as
\begin{equation}
\alpha_{\mathtt{ada}}^{(n)}=1-\Psi_\omega(t_n).
     \label{eq:ada}
\end{equation}
Similar to absorption, the term $\Psi_\omega(t_n)$ denotes the gap between $\SSE_{1},\SE_{2},\ldots,\SE_{K}$ at time $t_n$ and the desired value with weights $(\omega_1,\omega_2)$, where $\SSE_1^{\des}, \SE_k$ can be set to different values for various security application scenarios. Also, $\alpha^{(n)}_{\mathtt{ada}}$ can be achieved at an optimal value of 1.

\subsubsection{Recovery}In the part of \textit{recovery}, the timeliness of system recovery is evaluated by comparing the actual recovery time with the desired recovery time $T_d$. Here, $t_0$ represents the time of failure, and $t_n$ is the time at which the system recovers via its adaptation mechanism. The recovery is defined as
\begin{equation}
{\alpha^{(n)}_{\mathtt{rec}}} = \begin{cases}
1,&{\text{if}}\ t_n-t_0\le T_d \\ 
{\frac{T_d}{t_n-t_0},} &{\text{otherwise}} 
\end{cases},
\label{eq:rec}
\end{equation}
where the ideal result can be obtained when the adaptation mechanism allows the system to recover within the desired time (i.e., $t_n-t_0\le T_d$). Thus, a faster recovery yields a higher score.

The overall resilience metric comprises three components: absorption, adaptation, and recovery. The overall resilience score at iteration $n$ is defined as 
\begin{equation}
    \alpha^{(n)}=\lambda_1\alpha_{\mathtt{abs}}+\lambda_2\alpha^{(n)}_{\mathtt{ada}} +\lambda_3\alpha^{(n)}_{\mathtt{rec}},
    \label{overall}
\end{equation}
where the weighting coefficients $\lambda_1,\lambda_2,\lambda_3$ need to be non-negative and satisfy $\lambda_1+\lambda_2+\lambda_3=1$.
Originating in a single-criticality-level resilience scenario \cite{Reifert1}, our proposed secure resilience metric, $\alpha^{(n)}$, can effectively assess the resilience of the CF-mMIMO network across varying security requirements.


As discussed in \cite{Reifert1}, resilience metrics can be classified into anticipatory and reactionary actions based on the time dimension. Given the extensive prior work on robustness, we focus on reactionary mechanisms in this work. Accordingly, we set $\lambda_1 = 0$ so that $\alpha_{\mathtt{abs}}$ does not contribute to the final score. To efficiently meet resilience requirements, we adapt the standard SCA procedure by fixing the maximum number of iterations $N_{\max}$ in the main loop, rather than running the algorithm until full convergence. Each intermediate solution generated during these iterations is treated as a valid and deployable candidate. 

This enables explicit measurement of computation time for evaluating the recovery metric $\alpha_{\mathtt{rec}}$ weighted by $\lambda_3$ and allows selecting intermediate solutions that may outperform converged results in time-sensitive scenarios. Guided by the resilience priority weights $\lambda_i$, the framework calculates an overall resilience score via (\ref{overall}) after each iteration, reflecting both solution quality and computation time. After all iterations are completed, the algorithm selects the candidate solution with the highest score observed throughout the entire iteration record as the final output, rather than simply returning the last solution. Notably, when $\lambda_3=0$ (without considering recovery time) or when $T_d$ is sufficiently large (so that $\alpha_{\mathtt{rec}}\approx 1$), the solution coincides with the conventional SCA result that minimizes $\Psi_\omega$, and the proposed algorithm converges consistently to traditional optimization methods. The detailed steps of the proposed algorithm are provided in \textbf{Algorithm \ref{alg1}}.
\begin{algorithm}[!t]
    \caption{Secure Resilience-Aware SCA Optimization}
    \label{alg1}
    \begin{algorithmic} [1]   
        \STATE \textbf{Input:} $\qu^{(0)}, \boldsymbol{\tau}^{(0)}, \eta_1^{(0)}, \omega_1, \omega_2, \lambda_1, \lambda_2, \lambda_3, T_d, N_{\max}$;
        \STATE \textbf{Initialize:} $t_0,  \alpha_{\mathtt{best}}, \qu^*$
        \FOR{$n=1$ to $N_{\max}$}
        \STATE Record current time $t_n$;
        \STATE Obtain $(\qu^{(n)}, \boldsymbol{\tau}^{(n)}, \eta_1^{(n)}, \zeta_1^{(n)})$ by solving the SOCP problem (\ref{eq:SOCP}) using $(\qu^{(n-1)}, \boldsymbol{\tau}^{(n-1)}, \eta_1^{(n-1)})$;
        \STATE Calculate $\Psi_\omega$ by using $\qu^{(n)}$ in \eqref{obj_omega};
        \STATE Calculate adaptation metric $\alpha_{\mathtt{ada}}^{(n)}$ and recovery metric $\alpha_{\mathtt{rec}}^{(n)}$ by using $t_n$ in (\ref{eq:ada}) and (\ref{eq:rec});
        \STATE Calculate overall resilience $\alpha^{(n)}$ by using (\ref{overall});
        \IF{$\alpha^{(n)}> \alpha_{\mathtt{best}}$}      \STATE$\alpha_{\mathtt{best}}=\alpha^{(n)}, \qu^*=\qu^{(n)}$;
        \ENDIF
        \STATE Update  $(\qu^{(n+1)}, \boldsymbol{\tau}^{(n+1)}, \eta_1^{(n+1)}) =(\qu^{(n)}, \boldsymbol{\tau}^{(n)}, \eta_1^{(n)})$;
        \ENDFOR
        \STATE \textbf{Output:} The best-found resilient $\qu^{*}$.
    \end{algorithmic}
\end{algorithm}

\section{Numerical Results}
 \begin{figure*}[t]
    \centering
    \begin{subfigure}[t]{0.325\textwidth}
        \centering
        \includegraphics[width=\textwidth]{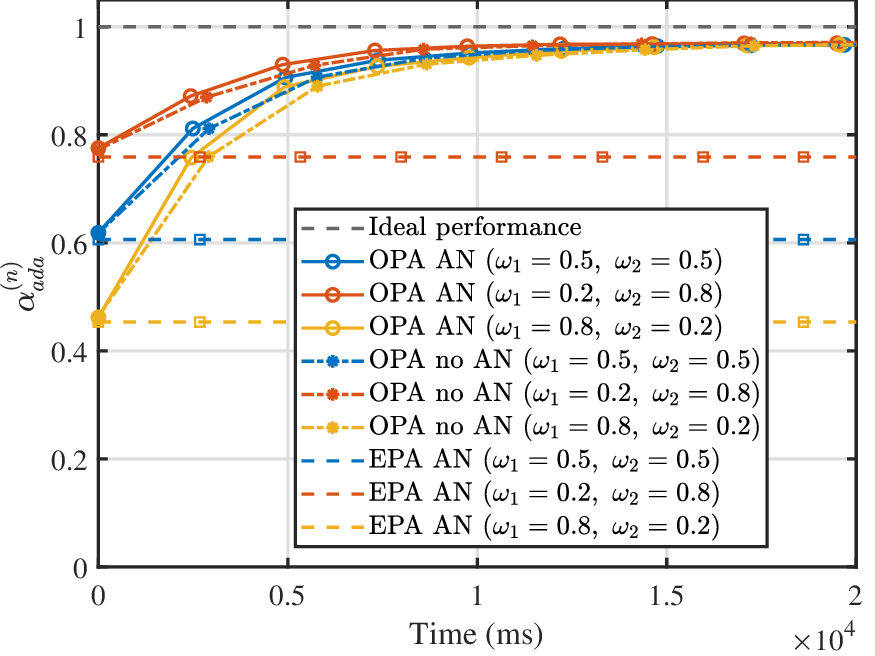}
        \caption{Convergence of $\alpha_{\mathtt{ada}}^{(n)}$ with different $(\omega_1,\omega_2)$ and $\lambda_1=0,\lambda_2=1,\lambda_3=0$.}
        \label{fig:SE_v_thickness}
    \end{subfigure}
    \hfill
    \begin{subfigure}[t]{0.325\textwidth}
        \centering
        \includegraphics[width=\textwidth]{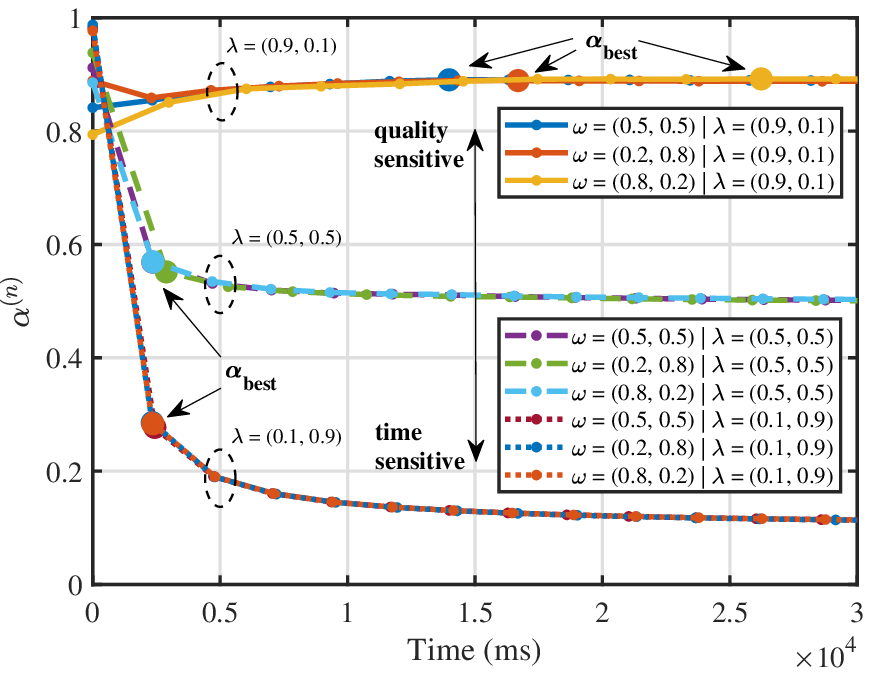}
        \caption{Different $(\omega_1,\omega_2)$ and $(\lambda_2,\lambda_3)$ after an outage occurs at $T_0=500$\,ms.}
        \label{fig:overall1}
    \end{subfigure}
    \hfill
    \begin{subfigure}[t]{0.325\textwidth}
        \centering
    \includegraphics[width=\textwidth]{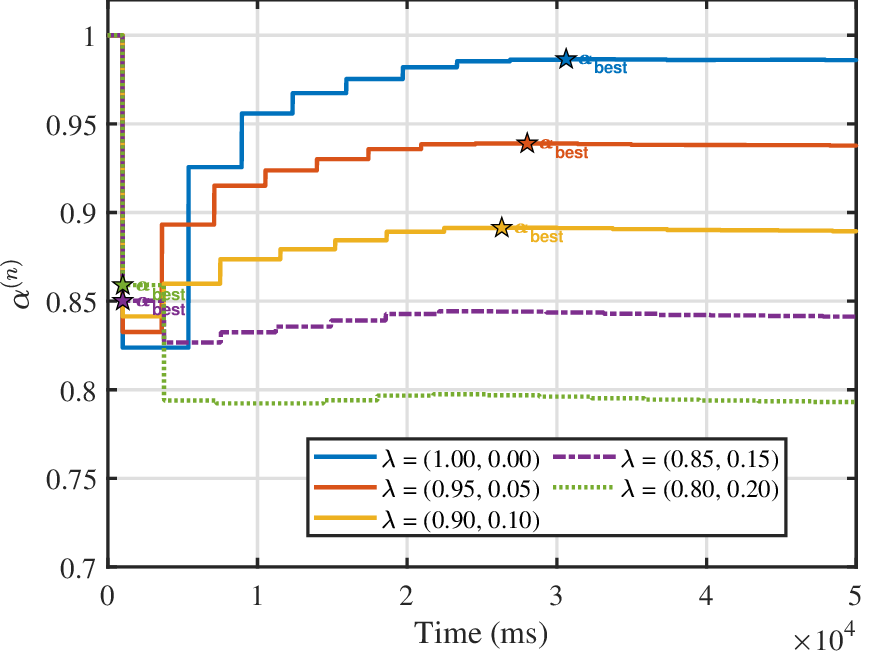}
        \caption{$\omega=(0.5,0.5)$ and different $(\lambda_2,\lambda_3)$ after an outage occurs at $T_0=1000$\,ms.}
        \label{fig:overall2}
    \end{subfigure}
    \vspace{-0.2em}
    \caption{The performance evaluation of the resilience framework: \textbf{Algorithm \ref{alg1}}.}
    \vspace{-1.2em}
    \label{fig:overall}
\end{figure*}
 In this section, we utilize numerical results to demonstrate the secrecy resilience of CF-mMIMO with different precoding schemes in scenarios involving an Eve. The users and APs are randomly distributed in a square of size $1\times1$\,$\text{km}^2$. The Eve is randomly placed in a circular area with a maximum radius of 0.1 km around $U_1$. Following \cite{yasseen1},\cite{LPZF}, the large-scale fading is modeled as $\beta_{l,\xi}=10^\frac{PL_{l,k}^d}{10}\times10^\frac{F_{l,k}}{10}$, where the first and second terms represent the path loss and shadowing effect, respectively. The path loss is given by $PL_{l,k}^d=-30.5-36.7\cdot\text{log}_{10}(\frac{d_{l,k}}{1\,(m)})$, where $d_{l,k}$ is the distance between the $l$-th AP and $U_k$. The shadowing effect follows the distribution $F_{l,k}\sim\mathcal{N}(0,4^2)$ (dB). In addition, we set $M=4, L=40, r=100$\,m, $P_{\max,l}=200$\,mW and $\tau_p=K=10$. The noise power is $-96$\,dBm and $T_d=500$\,ms. Finally, $\SSE^{\des}_1=3$\,bit/s/Hz, $\SE^{\des}_k=5$\,bit/s/Hz and $\SE^{\min}_k=0.1$\,bit/s/Hz.

We investigate the convergence of \textbf{Algorithm \ref{alg1}} in Fig.\,\ref{fig:overall}(a), which only considers the recovery quality (i.e., $\lambda_2=1$, $\lambda_3=0$) under different $\omega_1$ and $\omega_2$. To demonstrate the effectiveness of the proposed power allocation optimization and AN design, we compare it against two benchmark schemes: (i) the ``\textbf{OPA no AN}'' scheme, which applies optimal power allocation without AN, and (ii) the ``\textbf{EPA AN}'' scheme, which allocates equal power among the users while employing AN. The closer the value of $\alpha_{\mathtt{ada}}^{(n)}$ approaches 1, the nearer the system is to its desired performance ($\Psi=0$) (i.e., the higher the recovery quality). From Fig.\,\ref{fig:overall}(a), during the initial phase, curves with different values of $\omega$ all exhibit a relatively rapid recovery rate, which demonstrates the effectiveness of the proposed \textbf{Algorithm \ref{alg1}}. Specifically, the initial performance of each curve depends on the alignment between the power allocation strategy adopted at the initial feasible point (i.e., user power is evenly divided with no AN power) and the optimization objectives emphasized at different $\omega$ values. When the optimization objective focuses on security performance ($\omega_1=0.8$, $\omega_2=0.2$), the initial power allocation strategy has a lower initial performance due to the lack of power allocation for the AN. Subsequently, the algorithm performs power allocation for the AN, enabling the yellow curve to achieve a rapid performance improvement. Conversely, when the value of $\omega$ is set to align more closely with the initial strategy, the initial performance is higher. Ultimately, the curves converge to a similar high-performance level ($\alpha_{\mathtt{ada}}\approx 0.98$) around $1.7\times 10^4$\,ms.

We evaluate the overall resilience of \textbf{Algorithm \ref{alg1}} in Fig.\,\ref{fig:overall}(b) under different performance weights $\omega=(\omega_1,\,\omega_2)$ and quality-time score weights $\lambda=(\lambda_2,\lambda_3)$. Note that $\alpha_{\mathtt{ada}}^{(n)}$ typically reaches its optimal quality with increasing time, while $\alpha_{\mathtt{rec}}^{(n)}$ decreases over time. Consequently, the curves exhibit different trends and the long-term trend exhibits a negative correlation with $\lambda_2$. Specifically, when $\lambda=(0.9,0.1)$, the curve stabilizes near $0.9\times0.98$; stabilizes near $0.5\times0.98$ for $\lambda=(0.5,0.5)$, and approaches $0.1\times0.98$ for $\lambda=(0.1,0.9)$. In the initial stages, groups with larger recovery weights (e.g., $\lambda=(0.1,0.9)$) can commence with higher scores because $\alpha_{\mathtt{rec}}^{(n)}=1$ when $t_n-t_0\le T_d$. However, once $t_n> T_d$, the score declines sharply as $T_d/(t_n-t_0)$ diminishes. Conversely, when adaptation weights dominate (e.g., $\lambda=(0.9,0.1)$), the curve primarily follows the increasing trend of $\alpha_{\mathtt{ada}}^{(n)}$, rising steadily to high level.


In Fig.\,\ref{fig:overall}(c), the service priority weights $(\omega_1,\omega_2)$ are both fixed at 0.5. The figure shows the selection and update process of the best-found resilient $\alpha_{\mathrm{best}}$, as defined in \textbf{Algorithm \ref{alg1}}. Specifically, when the system prioritizes recovery quality (e.g., $\lambda=(1,0)$ and $\lambda=(0.95,0.05)$, its instantaneous resilience value $\alpha^{(n)}(t)$ continuously optimizes and increases with iterations. Consequently, these quality-sensitive curves rise when spending more time to obtain high-quality solutions. Conversely, when the resilience weights for recovery time increases (e.g., $\lambda=(0.85,0.15)$ and $\lambda=(0.8,0.2)$), the system tends toward a more “short-sighted” strategy. Since the recovery time component $\alpha_\mathtt{rec}^{(n)}$ holds a high score in the early stages, these curves are likely to achieve a locally optimal $\alpha^{(n)}$ value at an earlier time and lock it as $\alpha_\mathrm{best}$. Thereafter, although the algorithm's recovery quality $\alpha_\mathtt{ada}^{(n)}$ continues to improve, the rapid decline of the time penalty term prevents subsequent instantaneous $\alpha^{(n)}(t)$ values from surpassing this early peak. Consequently, the term $\alpha_\mathrm{best}$ of these curves is rapidly determined early on and remains fixed, and their final performance upper bound is significantly lower than that of quality-sensitive strategies.

\vspace{-0.2em}
\section{Conclusion}
In this work, we studied the security resilience of a CF-mMIMO system under active eavesdropping by proposing a novel resilience-aware optimization framework. We formulated a power allocation problem governed by service priorities $(\omega_i)$ that minimizes a performance gap, balancing security against QoS. The proposed resilience-aware SCA algorithm then utilizes a set of resilience priorities $(\lambda_i)$ to make a final decision by trading off the solution quality against recovery time.

\bibliographystyle{IEEEtran}
\bibliography{IEEEabrv,Reference}

@STRING{IEEE_J_VT         = "{IEEE} Trans. Veh. Technol."}

@STRING{IEEE_J_SP         = "{IEEE} Trans. Signal Process."}

@STRING{IEEE_J_COML       = "{IEEE} Commun. Lett."}

@STRING{IEEE_J_COM        = "{IEEE} Trans. Commun."}

@STRING{IEEE_J_WCOM       = "{IEEE} Trans. Wireless Commun."}

@STRING{IEEE_J_IFS        = "{IEEE} Trans. Inf. Forensics Security"}

@STRING{IEEE_J_MC         = "{IEEE} Trans. Mobile Comput."}

@STRING{IEEE_J_PROC       = "Proc. {IEEE}"}

@STRING{IEEE_M_COM        = "{IEEE} Commun. Mag."}

@STRING{IEEE_O_CSTO        = "{IEEE} Commun. Surveys Tuts."}

@STRING{IEEE_OJVT         = "{IEEE} Open J. Veh. Technol."}

@STRING{IEEE_OJCS         = "{IEEE} Open  J. Commun. Society"}

@STRING{IEEE_O_CSTO       = "{IEEE} Commun. Surveys Tuts."}

@ARTICLE{Hoang:TWCOM:2018,
  author={Hoang, Tiep M. and Ngo, Hien Quoc and Duong, Trung Q. and Tuan, Hoang Duong and Marshall, Alan},
  journal=IEEE_J_COM, 
  title={Cell-Free Massive {MIMO} Networks: Optimal Power Control Against Active Eavesdropping}, 
  year={2018},
  volume={66},
  number={10},
  pages={4724-4737},
  doi={10.1109/TCOMM.2018.2837132},
  ISSN={1558-0857},
  month={Oct.},}

@ARTICLE{mohammadi2024next,
  author={Mohammadi, Mohammadali and Mobini, Zahra and Ngo, Hien Quoc and Matthaiou, Michail},
  journal=IEEE_J_PROC, 
  title={Next-Generation Multiple Access With Cell-Free Massive {MIMO}}, 
  year={2024},
  volume={112},
  number={9},
month = {Sept.},
  pages={1372-1420},
  doi={10.1109/JPROC.2024.3451372}}

@article{matthaiou2021road,
  title={The road to {6G}: Ten physical layer challenges for communications engineers},
  author={Matthaiou, Michail and Yurduseven, Okan and Ngo, Hien Quoc and Morales-Jimenez, David and Cotton, Simon L and Fusco, Vincent F},
  journal=IEEE_M_COM,
  volume={59},
  number={1},
  pages={64-69},
  year={2021},
  month={Jan.},
  publisher={IEEE}
}

@ARTICLE{hien,
  author={Ngo, Hien Quoc and Ashikhmin, Alexei and Yang, Hong and Larsson, Erik G. and Marzetta, Thomas L.},
  journal=IEEE_J_WCOM, 
  title={Cell-Free Massive {MIMO} Versus Small Cells}, 
  year={2017},
  volume={16},
  number={3},
  pages={1834-1850},
  doi={10.1109/TWC.2017.2655515},month={Mar.}}

@ARTICLE{Junbin,
  author={Yu, Junbin and Zhao, Sai and Tian, Maoxin and Li, Quanzhong and Tang, Dong},
  journal=IEEE_J_COML,
  title={Joint Transmit Antenna Selection and Beamforming Design for Millimeter Wave Massive {MIMO} {NOMA} Secure Networks}, 
month = {Oct.},
  year={2024},
  volume={28},
  number={10},
  pages={2407-2411},
  doi={10.1109/LCOMM.2024.3438876}}

@ARTICLE{yasseen2,
  author={Atiya, Yasseen Sadoon and Mobini, Zahra and Ngo, Hien Quoc and Matthaiou, Michail},
  journal=IEEE_OJCS, 
  title={Cell-Free Massive {MIMO} With Multiple Active Eavesdroppers}, 
month= {Jan.},
  year={2025},
  volume={6},
  number={},
  pages={1859-1872},
  doi={10.1109/OJCOMS.2025.3534640}}

@ARTICLE{yasseen1,
  author={Atiya, Yasseen Sadoon and Mobini, Zahra and Ngo, Hien Quoc and Matthaiou, Michail},
  journal=IEEE_J_WCOM,
  title={Secure Transmission in Cell-Free Massive {MIMO} Under Active Eavesdropping}, 
month = {Dec.},
  year={2024},
  volume={23},
  number={12},
  pages={18036-18052},
  doi={10.1109/TWC.2024.3459628}}

@ARTICLE{PLS6G,
  author={Mitev, Miroslav and Chorti, Arsenia and Poor, H. Vincent and Fettweis, Gerhard P.},
  journal=IEEE_OJVT, 
  title={What Physical Layer Security Can Do for {6G} Security}, 
month= {Feb.},
  year={2023},
  volume={4},
  number={},
  pages={375-388},
  doi={10.1109/OJVT.2023.3245071}}

@ARTICLE{Reifert1,
  author={Reifert, Robert-Jeron and Roth, Stefan and Ahmad, Alaa Alameer and Sezgin, Aydin},
  journal=IEEE_J_VT, 
  title={Comeback {Kid}: Resilience for Mixed-Critical Wireless Network Resource Management}, 
month={Dec.},
  year={2023},
  volume={72},
  number={12},
  pages={16177-16194},
  doi={10.1109/TVT.2023.3296977}}

@ARTICLE{LPZF,
  author={Interdonato, Giovanni and Karlsson, Marcus and Björnson, Emil and Larsson, Erik G.},
  journal=IEEE_J_WCOM, 
  title={Local Partial Zero-Forcing Precoding for Cell-Free Massive {MIMO}},
month= {Jul.},
  year={2020},
  volume={19},
  number={7},
  pages={4758-4774},
  doi={10.1109/TWC.2020.2987027}}

@ARTICLE{Femenias,
  author={Femenias, Guillem and Riera-Palou, Felip},
  journal=IEEE_J_MC,
  title={From Cells to Freedom: {6G}'s Evolutionary Shift With Cell-Free Massive {MIMO}}, 
  year={2025},month={Feb.},
  volume={24},
  number={2},
  pages={812-829},
  doi={10.1109/TMC.2024.3468003}}

@ARTICLE{Tubail,
  author={Tubail, Deeb Assad and Alsmadi, Malek and Ikki, Salama},
  journal=IEEE_J_IFS, 
  title={Physical Layer Security in Downlink of Cell-Free Massive {MIMO} With Imperfect {CSI}}, 
  year={2023},month={May},
  volume={18},
  number={},
  pages={2945-2960},
  doi={10.1109/TIFS.2023.3272769}}

@ARTICLE{AN,
  author={Tsai, Shang-Ho and Poor, H. Vincent},
  journal=IEEE_J_SP, 
  title={Power Allocation for Artificial-Noise Secure {MIMO} Precoding Systems}, 
month= {Jul.},
  year={2014},
  volume={62},
  number={13},
  pages={3479-3493},
  doi={10.1109/TSP.2014.2329273}}

@ARTICLE{ASAN,
  author={Niu, Hong and Xiao, Yue and Lei, Xia and Chen, Jiangong and Xiao, Zhihan and Li, Mao and Yuen, Chau},
  journal=IEEE_O_CSTO,
  title={A Survey on Artificial Noise for Physical Layer Security: opportunities, Technologies, Guidelines, Advances, and Trends}, 
  year={2026},
  volume={28},
  number={},
  pages={341-381},
  doi={10.1109/COMST.2025.3610758}}

\end{document}